\documentclass[referee]{aa}
\usepackage{txfonts}
\usepackage{graphicx}
\usepackage[longnamesfirst]{natbib}
\bibpunct{(}{)}{;}{a}{}{,}
\begin{document} 

   \title{Dust emissivity in the Submm/Mm}
   \subtitle{SCUBA and SIMBA observations of Barnard~68}
 
   \author{
       S. Bianchi\inst{1}
       \and
       J. Gon\c{c}alves\inst{2,3}
       \and
       M. Albrecht\inst{4}
       \and
       P. Caselli\inst{2}
       \and
       R. Chini\inst{4}
       \and
       D. Galli\inst{2}
       \and
       M. Walmsley\inst{2}
   }
 
   \offprints{S. Bianchi, sbianchi@arcetri.astro.it}
 
   \institute{ 
   CNR-Istituto di Radioastronomia -- Sezione di Firenze, 
       Largo E. Fermi 5, I-50125 Firenze, Italy
   \and
   INAF-Osservatorio Astrofisico di Arcetri, 
       Largo E. Fermi 5, I-50125 Firenze, Italy
   \and
   Centro de Astronomia e Astrof\'{\i}sica da Universidade de Lisboa,
       Tapada da Ajuda, 1349-018 Lisboa, Portugal
   \and
   Astronomisches Institut der Ruhr-Universit\"at Bochum, 
       Universit\"atsstr.  150, D--44780 Bochum, Germany
   } 
 
   \date{Received  / Accepted } 
 
\abstract{
We have observed the dark cloud Barnard~68 with SCUBA at 850 $\mu$m and with 
SIMBA at 1.2 mm. The submillimetre and millimetre dust emission correlate well
with the extinction map of \citet{AlvesNature2001}.  
The A$_\mathrm{V}$/850 $\mu$m correlation is clearly not linear and suggests 
lower temperatures for the dust in the inner core of the cloud. Assuming a 
model for the temperature gradient, we derive the cloud-averaged  dust emissivities 
(normalised to the V-Band extinction efficiency) at 850 $\mu$m and 1.2 mm. 
We find $\kappa_{850\mu\mathrm{m}}/\kappa_\mathrm{V}=4.0\pm 1.0\cdot 
10^{-5}$ and $\kappa_{1.2\mathrm{mm}}/\kappa_\mathrm{V}=9.0\pm 3.0\cdot
10^{-6}$. These values are compared with other determinations in 
this wavelength regime and with expectations for models of diffuse dust and 
grain growth in dense clouds.
\keywords{ Radiation mechanisms: thermal  -- dust, extinction -- ISM: clouds, 
individual objects: Barnard 68 -- Submillimeter -- Radio continuum: ISM }
}

\maketitle 

%\shortcites{}
\shortcites{AltonSub1999}
\shortcites{JamesMNRAS2002}
\shortcites{KramerA&A1998}
\shortcites{LadaApJ1994}
\shortcites{HollandMNRASprep1998}
\shortcites{NymanMsngr2001}
\shortcites{BoulangerA&A1996}
\shortcites{BerginApJL2002}
\shortcites{HotzelA&A2002a}
\shortcites{MennellaApJ1998}
\shortcites{AveryApJ1987}
\shortcites{KramerA&A2002}
%__________________________________________________________________
\section{Introduction}

Despite being a fundamental parameter in Far Infrared, Submillimetre and 
Millimetre astronomy, few measurements of the dust emissivity\footnote{The 
emissivity proper is the emission [absorption] cross section normalized to 
the geometrical cross section of a dust grain. Other authors prefer the 
cross section normalized to the mass of the grain. Both quantities are the 
same when normalised to the analogous quantity for V-band extinction.} are 
available \citep[for a review, see][]{AltonSub1999,JamesMNRAS2002}.
The submm/mm dust emissivity is particularly important for star formation
%studies. Since molecules are known to deplete in the interior of prestellar 
studies. Since molecules are known to deplete inside prestellar 
cores \citep{BerginApJL2002}, dust emission may represent the best tracer 
of the gas density distribution just prior to the onset of gravitational 
collapse. Thus measurements in the submm/mm define 
the initial conditions from which a core collapses to form a star.
Alternatively, the density distribution can be mapped through extinction,
by measuring the near-infrared colour excess towards giant stars in the 
background of a cloud \citep{LadaApJ1994}.
High resolution and S/N maps can be obtained for object in the
foreground of dense stellar fields \citep{AlvesNature2001}. 

This is the case for the dark cloud Barnard~68, a starless globule seen 
in the foreground of the Galactic Bulge. \citet{AlvesNature2001} have
produced a high resolution extinction map of the cloud, measuring the 
H-K colour excess of nearly 4000 stars in its background. 
In this paper, we compare the extinction map with observations of 
submm/mm dust emission of similar resolution.  Observations are described in
Sect.~\ref{obse}. In Sect.~\ref{ana} we derive the dust emissivity from
the correlation between emission and extinction (in a way similar to
\citealt{KramerA&A1998,KramerA&A2002}). We will also adopt a temperature 
gradient within the cloud, which was derived from a model of
dust heating. Finally, the derived emissivities are compared with other
estimates from literature in Sect.~\ref{conc}.

%__________________________________________________________________
\section{Submm/mm Observations}
\label{obse}

\begin{figure*}[!t]
\sidecaption
\resizebox{12cm}{!}{ 
\includegraphics{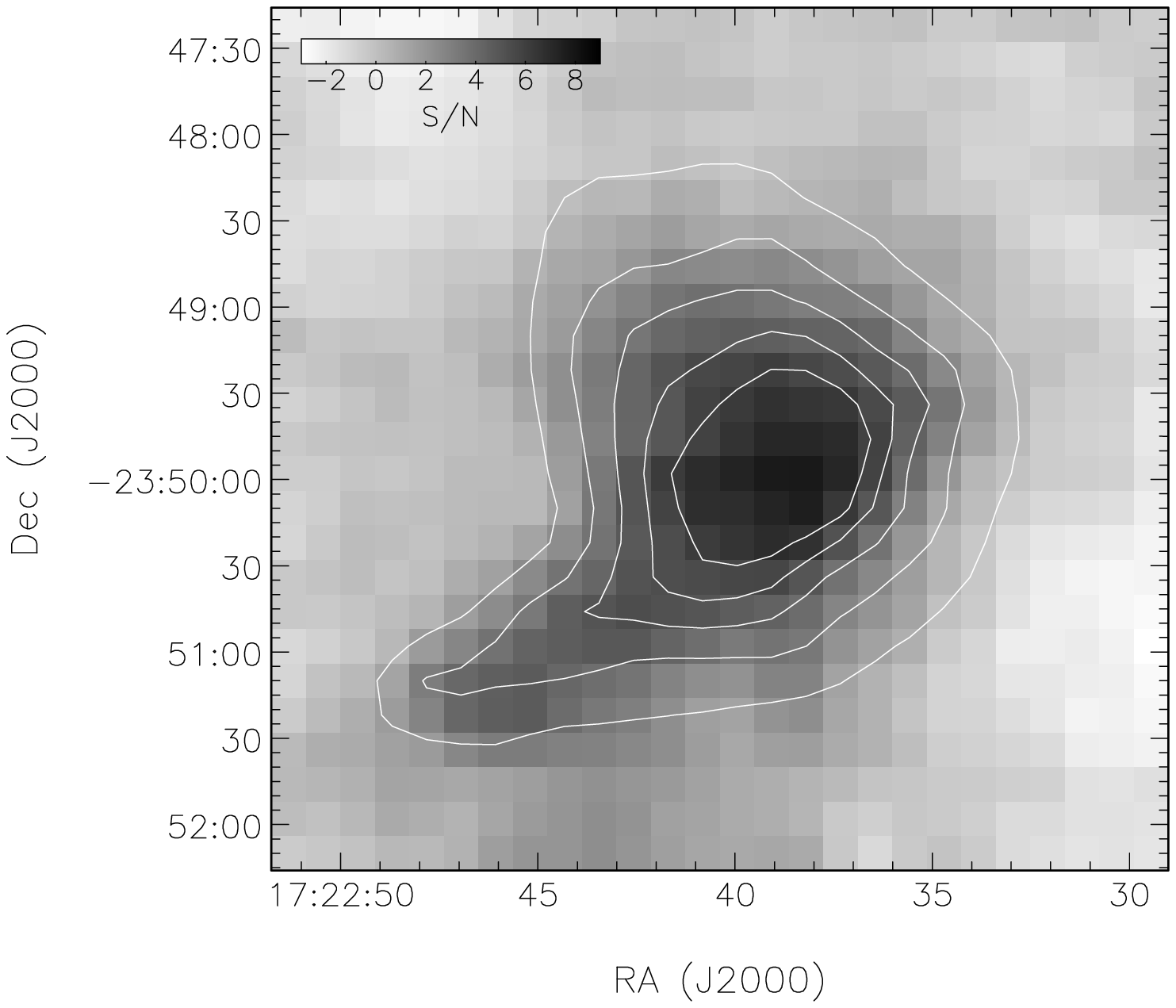}
\includegraphics{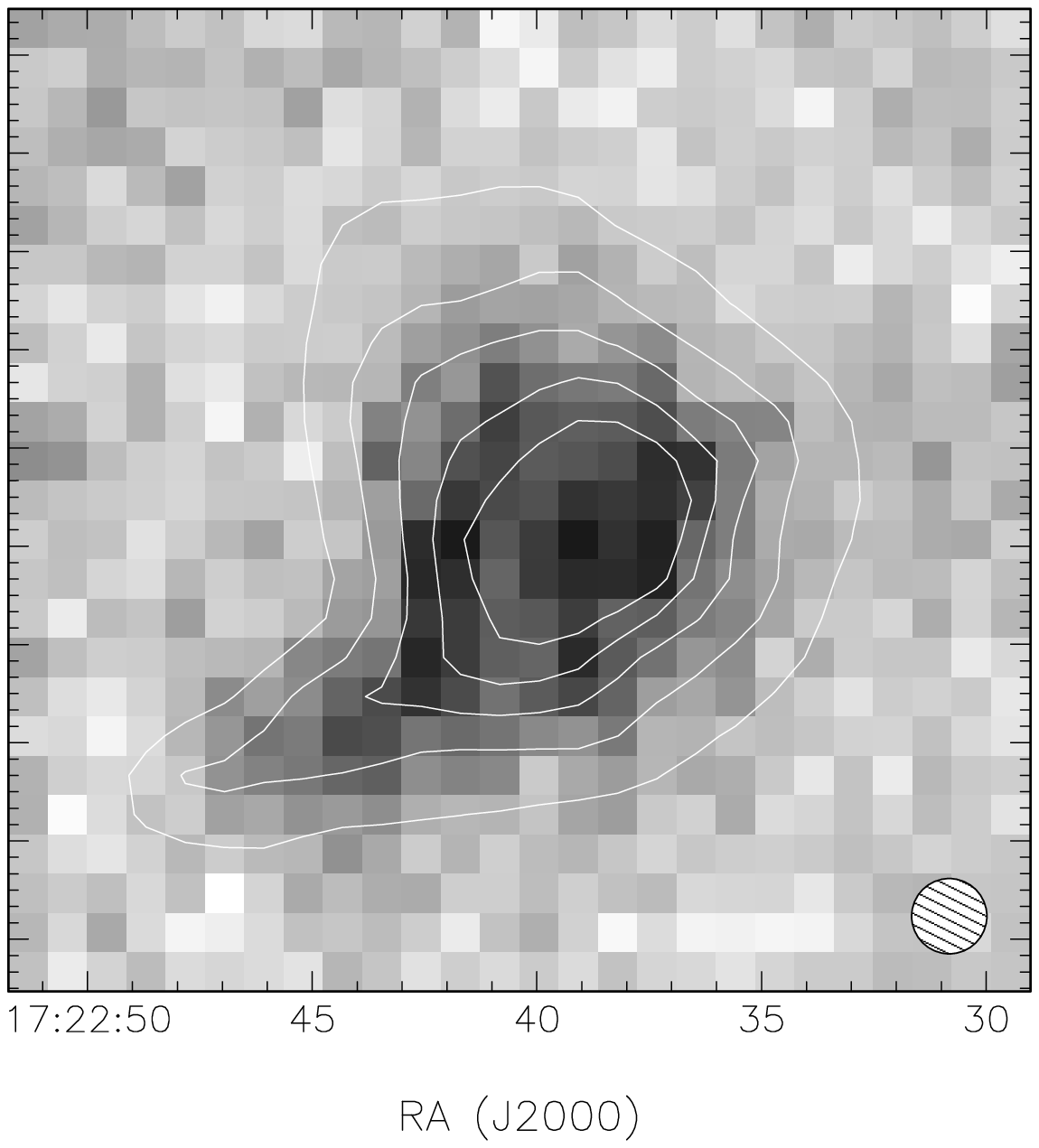} 
}
\caption{SCUBA map at 850 $\mu$m (left) and SIMBA map at 1.2 mm (right) 
of Barnard~68, with superimposed $A_\mathrm{V}$ contours. The $A_\mathrm{V}$ 
and SCUBA maps have been smoothed to match the SIMBA resolution 
(FWHM=24\arcsec; the beamsize is shown on the right image). All images have 
been resampled to a pixel size of 12\arcsec. The field of view is 5\arcmin
$\times$5\arcmin. $A_\mathrm{V}$ contours start at 4 mag and are spaced by 
4 mag. For both images, the grayscale in units of S/N. }
\label{allmaps}
\end{figure*}

Barnard~68 was observed with the Submillimetre Common User Bolometer
Array (SCUBA) at the JCMT \citep{HollandMNRASprep1998} and with
the SEST Imaging Bolometer Array \citep[SIMBA;][]{NymanMsngr2001}.
Both instruments have hexagonal arrays of 37 bolometers: SCUBA can
observe in the submm at 850 $\mu$m with FWHM=14.5\arcsec\ 
while SIMBA operates at 1.2 mm and has a beamsize of 24\arcsec.

SCUBA observed an area of 5\arcmin$\times$5\arcmin\ around Barnard~68 
in scan-map mode, chopping within the field of view. The resultant image is 
convolved with the chop function, which is removed by means of Fourier 
Transform analysis. The Emerson II technique is used to minimise the
noise \citep{JennessProc1998}: six scans of the field are made with
different chop throws (20\arcsec, 30\arcsec\ and 65\arcsec\ in RA and
Dec). In total, we coadded eight sets of scans, six observed in March/June 
2002 and two in July 1998 (the latter retrieved from the SCUBA Archive).
Standard data reduction was performed with the dedicated package SURF. After 
flat-fielding and masking of noisy bolometers, a baseline was removed 
from each bolometer. Maps were corrected for atmospheric extinction
(the zenith optical depth was $\tau_{850}$=0.12,0.15,0.3 in the 1998 run
and in the March/June 2002 runs, respectively) and for correlated sky 
noise. Calibration was obtained from scan-maps of Uranus.
The final map has a sky noise of 20 mJy beam$^{-1}$ (1-$\sigma$), equivalent 
to 3.5 MJy sr$^{-1}$ for the SCUBA beamsize.

SCUBA scan-maps suffer from large scale undulations in the background of the 
source, which are due both to the poor sampling of small spatial frequencies 
and to uncertainties in the baseline removal \citep[see, e.g.][]{VisserAJ2002}.
On our image, the undulations show up as a negative background around the 
source, of the order of the sky noise. We subtracted a mean background 
estimated on several sky apertures. The results of the next section are not 
significantly changed by the subtraction.
The integrated flux at 850 $\mu$m inside the 2-$\sigma$ isophote is
$F_\mathrm{850\mu m}$=4.0$\pm$1.0 Jy. SCUBA also took maps with the  450 
$\mu$m array. The signal at this wavelength is poor, because of the background 
fluctuations and of the larger sky opacity ($\tau_{450}\approx 0.5$ for the best 
data). The peak emission is detected at 3-$\sigma$ level (1-$\sigma$ = 200 mJy 
beam$^{-1}$; beamsize FWHM=10\arcsec)
and it is not possible to conduct the surface brightness analysis described 
in Sect.~\ref{ana}.  Nevertheless, we could estimate the integrated 
flux, which is $F_\mathrm{450\mu m}$= 13$\pm$4 Jy.

SIMBA observations were taken in fast scanning mode, without using a 
wobbling secondary mirror \citep{NymanMsngr2001}. The bolometer array scans 
the observed area with a speed of 80\arcsec\ s$^{-1}$ in azimuth and a 
scan-to-scan separation of 8\arcsec\ in elevation. Spatial frequencies 
are converted into temporal frequencies and the sky signal is filtered out 
with a low-cut filter. Each map of Barnard~68 covers 
$600\arcsec\times 312\arcsec$ in azimuth and elevation, and takes 7
minutes to complete. Barnard~68 was observed during June and October 2001.
A total of 118 maps were coadded together, for a total integration time
of 13.8 hours. The mean zenith optical depth is $\tau_{1.2}$=0.16.

Data were reduced with the software package MOPSI by R.~Zylka. After 
deconvolution according to the frequency passband, subtracting a baseline 
to each scan, correcting for sky opacity and gain elevation the correlated 
sky noise fluctuations and artifacts due to electronics were removed.  
All maps of Barnard~68 were rebinned in a single image, with pixel 
size 8\arcsec.  In an iterative way, a preliminary image was used as source 
model to improve the removal of correlated noise. Uranus maps were used to 
calibrate the data. Night-to-night fluctuations suggest a calibration 
uncertainty of 20\% (1-$\sigma$). The final SIMBA image of Barnard~68 has 
a residual noise of 5.5 mJy beam$^{-1}$ (1-$\sigma$), equivalent to 0.36 MJy 
sr$^{-1}$ for the SIMBA resolution. The integrated flux is 
$F_\mathrm{1.2 mm}$=0.7$\pm$0.2 Jy

The emission is compared to the extinction map of \citet{AlvesNature2001} 
(the measured NIR colour excess has been converted to $A_\mathrm{V}$ using a 
standard reddening law; \citealt{RiekeApJ1985}). All images have been smoothed 
to the SIMBA resolution, registered together (pointing accuracy is
better than a few arcsec for the submm/mm observations) and resampled to a 
common pixel size of 12\arcsec. Fig.~\ref{allmaps} shows the final 
SCUBA (left) and SIMBA (right) images, with superimposed $A_\mathrm{V}$ 
contours.

%__________________________________________________________________
\section{Analysis}
\label{ana}

\begin{figure*}[!t]
\sidecaption
\centering
\includegraphics[width=12cm]{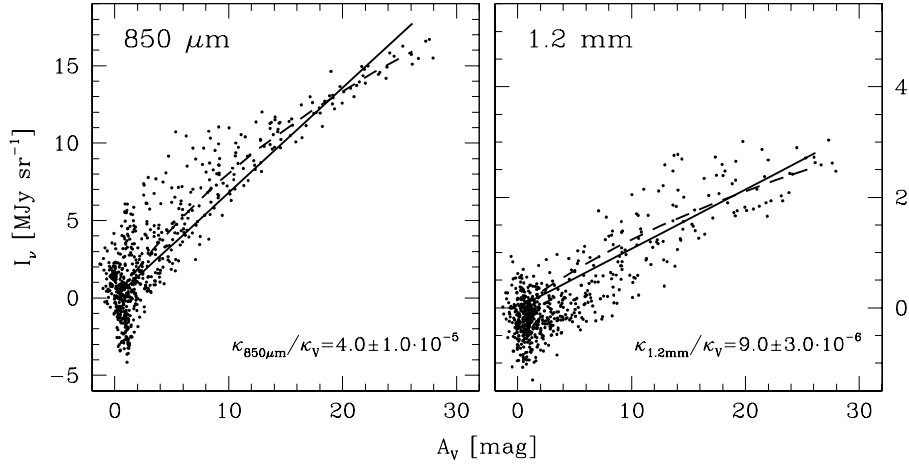} 
\caption{Pixel to pixel correlation between dust emission and
$A_\mathrm{V}$ extinction, at 850 $\mu$m (left) and 1.2 mm (right).
The solid line is the isothermal fit, assuming a temperature
$T_\mathrm{iso}$=12K, while the dashed line and the emissivity values
refer to the temperature gradient model with external temperature 
$T_\mathrm{ext}$=14K (see text for details). All high signal-to-noise 
points with $A_\mathrm{V}>10$ have been used for the fit.
}
\label{fitem}
\end{figure*}

The morphology of the extinction and emission maps in Fig.~\ref{allmaps} 
is quite similar. Dust emission with $S/N$$>$3 traces regions with 
extinction $A_\mathrm{V}$$>$6 and $A_\mathrm{V}$$>$10, in the SCUBA and SIMBA 
images, respectively. In analogy with extinction, emission comes from a 
round region with a south east tail. At 1.2 mm, there are hints for a
secondary emission peak. In Fig.~\ref{fitem} we show the 
pixel-to-pixel correlation of the emission maps with $A_\mathrm{V}$. As already 
seen in Fig.~\ref{allmaps}, there is a clear correlation between emission and 
extinction. For regions with $A_\mathrm{V}$$>$10 the scatter in the correlation 
with the submm data is of the order of the sky noise (1.7 MJy sr$^{-1}$ 
for the smoothed image). For $A_\mathrm{V}$$<$10 the scatter increases because 
of the (possibly hotter) south-east tail. The pixels clearly belonging to
the tail can be seen in Fig.~\ref{fitem}  around $A_\mathrm{V}$=8 and 
$I_\nu$=10 MJy sr$^{-1}$.  The scatter at 1.2 mm is 
more uniform and slightly larger than the sky noise, because of the
complex central morphology. A forthcoming paper will be devoted to the 
analysis of the central peaks and of the south-east tail.

For optically thin radiation (as for the submm/mm emission of Barnard 68)
we can write
\begin{equation}
I_\nu=\frac{\kappa_\nu}{\kappa_\mathrm{V}} \times \frac{A_\mathrm{V}}{1.086}
\times\frac{\int \rho \;\;B_\nu(T)\;\; dl}{\int \rho\;\; dl},
\label{emifind}
\end{equation}
where $T$ and $\rho$ are the temperature and density of dust grains,
functions of the position within the cloud, $B_\nu$ is the Planck function and 
the integral extends along the line of sight through the dust. In the 
isothermal case, the last term on the right reduces to $B_\nu(T)$.
Otherwise, a knowledge of the dust temperature and density distribution is 
needed.

\begin{figure}[b]
\centering
\includegraphics[width=7.0cm]{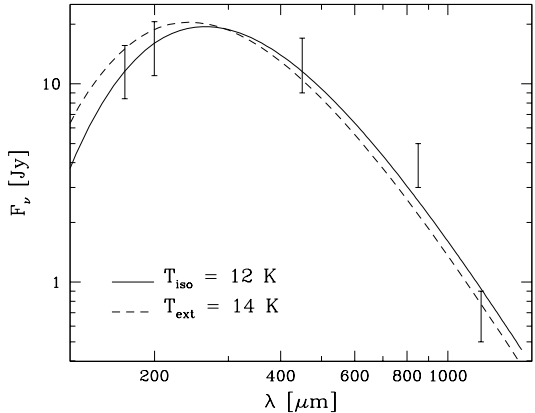} 
\caption{SED of Barnard 68. The solid line is the isothermal model with 
$T_\mathrm{iso}=12$ K. The dashed line is for the model with a temperature 
gradient, with $T_\mathrm{ext}=14$ K. Both cases assume $\beta=1.7$. 
}
\label{sed}
\end{figure}

In Fig.~\ref{sed} we show the Spectral Energy Distribution (SED) of Barnard 68.
Our integrated fluxes at 850 $\mu$m and 1.2 mm are similar to those presented
by \citet{WardThompsonMNRAS2002}, from which we took the ISOPHOT fluxes at 
$170\mu m$ and $200\mu m$. We fitted a isothermal modified blackbody to the 
datapoints assuming a value for $\beta$, the emissivity spectral index 
($\kappa_\nu \propto \nu^{\beta}$).  For $\beta$=1.5-2 \citep{DunneMNRAS2001}, 
we find acceptable fits for temperatures $T_\mathrm{iso}=$11-13 K. We adopt 
$T_\mathrm{iso}=12\pm 2$ K (1-$\sigma$) for the isothermal case, 
incorporating in the broad error the large uncertainties on $\beta$. 
The solid line in Fig.~\ref{sed} is an indicative SED for $\beta$=1.7 and 
$T_\mathrm{iso}$=12 K. Similar temperature have been previously obtained for
Barnard~68 \citep{WardThompsonMNRAS2002,HotzelA&A2002a}. Using $T_\mathrm{iso}$,
we fitted Eq.~\ref{emifind} to the correlations of Fig.~\ref{fitem} and
derived a single cloud-averaged value for the submm and mm emissivities.
We obtained
$\kappa_{850\mu\mathrm{m}}/\kappa_\mathrm{V}=3.5\pm 1.0\cdot 10^{-5}$
and $\kappa_{1.2\mathrm{mm}}/\kappa_\mathrm{V}=9.0\pm 3.0\cdot 10^{-6}$.
Errors were estimated with a bootstrap technique and are dominated by 
the uncertainty on calibration and on temperature.  

The fitted correlations are shown as solid lines in Fig.~\ref{fitem}.
While the fit is acceptable at 1.2 mm, a simple linear relation is unable 
to reproduce well the correlation observed at 850 $\mu$m for 
$A_\mathrm{V}>20$ since the $I_\nu/A_\mathrm{V}$ ratio decreases
for increasing $A_\mathrm{V}$. This is expected if the temperature in the 
cloud core is lower than in the external part, as a result of the dust 
shielding of the external radiation field \citep{ZucconiA&A2001}. Such
a trend has been observed in other clouds 
\citep[see, e.g.\ ][]{KramerA&A1998}. A similar behaviour could also be 
explained with a isothermal dust distribution, if the dust emissivity is 
lower in denser regions. However, models and observations suggest that
emissivities increase in dense cores \citep{OssenkopfA&A1994,KramerA&A2002}.

We derived the temperature gradient inside the cloud from an improved
version of the \citeauthor{ZucconiA&A2001} model \citetext{Gon\c{c}alves
et al., in preparation}. As in \citet{AlvesNature2001}, the adopted
density distribution is that for a Bonnor-Ebert sphere, a pressure confined 
isothermal sphere in hydrostatic equilibrium. Dust is heated by a
local Interstellar Radiation Field \citep{GalliA&A2002}. The internal
dust absorption is fixed by the measured extinction and the assumed
absorption law $\kappa_\nu/\kappa_\mathrm{V}$, for which we used
the tabulated values given by \citet{OssenkopfA&A1994} for different
models of dust coating/coagulation in dense clouds. While the 
absolute value of the temperature depends on the adopted dust model (for
$\lambda>30 \mu$m, $\kappa_\nu/\kappa_\mathrm{V}$ increases going from
bare grains to grains with ice coating up to grains that
undergo coagulation), the temperature gradient is found to be relatively
independent of grain characteristics, with the temperature at the 
truncation radius $T_\mathrm{ext}$ \citep[$0.06$~pc;][]{AlvesNature2001} 
about 1.5 times higher then the core temperature. This is because the 
absorption law does not change significantly with the dust model at shorter 
wavelengths, where most of the absorption occurs. In the following, we 
adopt the model radial gradient and derive $T_\mathrm{ext}$ from the data.

For the assumed density and temperature distributions, we simulated
emission maps with 24\arcsec\ resolution and 12\arcsec\ pixels, in 
analogy with the data. The integrated SED is shown with a dashed line in
Fig.~\ref{sed}. Again, a broad range of best fit temperatures is
obtained, $T_\mathrm{ext}$=12-15 K. We use $T_\mathrm{ext}=14\pm 2$ K
(1-$\sigma$). Finally, the simulated $A_\mathrm{V}$ vs $I_\nu$
correlation of  Eqn.~\ref{emifind} is fitted to the observed data to
derive $\kappa_\nu/\kappa_\mathrm{V}$. We obtained\footnote{Equivalent to
$\kappa_{850\mu\mathrm{m}}=1.5 \pm 0.4\; \mathrm{cm}^2\; \mathrm{g}^{-1}$ and
$\kappa_{1.2\mathrm{mm}}=0.35 \pm 0.1\; \mathrm{cm}^2\; \mathrm{g}^{-1}$
\citep[assuming grain radius 0.1$\mu$m, grain density 3 g cm$^{-3}$ and
V-band extinction efficiency 1.5;][]{HildebrandQJRAS1983}.}
\begin{eqnarray}
\kappa_{850\mu\mathrm{m}}/\kappa_\mathrm{V} &= & 4.0 \pm 1.0 \cdot 10^{-5} \nonumber\\
\kappa_{1.2\mathrm{mm}}/\kappa_\mathrm{V}   &= & 9.0 \pm 3.0 \cdot 10^{-6}. \nonumber
\end{eqnarray}
These correlations are also shown in Fig.~\ref{fitem} with dashed lines. 
Although the fit to the submm data has improved, the emissivities are
nearly the same as for the isothermal case.

\section{Discussion}
\label{conc}

\begin{figure}[!t]
\centering
\includegraphics[width=7.5cm]{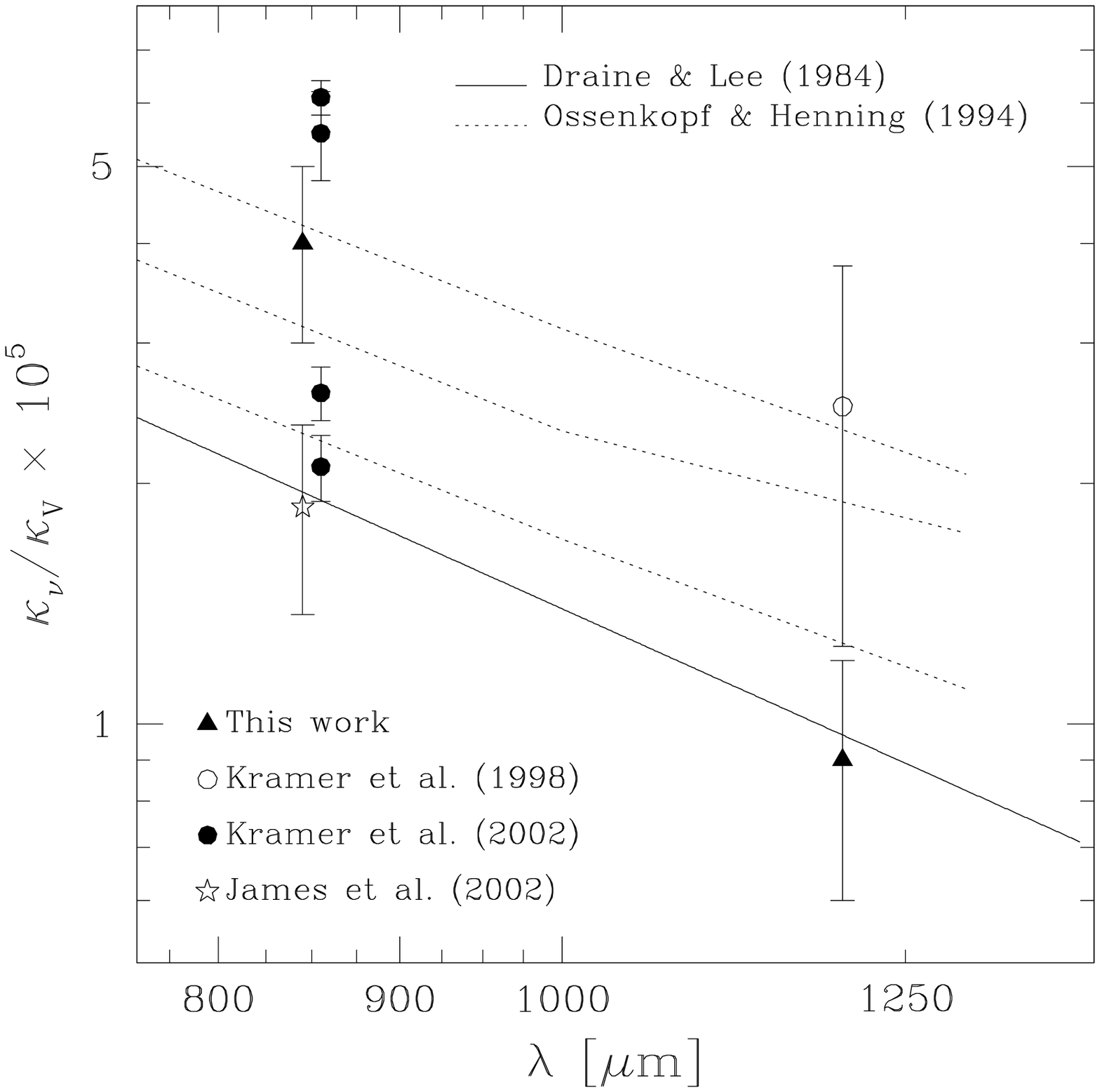}
\caption{Submm/mm emissivities compared with other literature estimates 
(details in the text). A standard ratio $\kappa_\mathrm{K}/\kappa_\mathrm{V}$ 
\citep{RiekeApJ1985} has been used to 
derive $\kappa_\nu/\kappa_\mathrm{V}$ from \citet{OssenkopfA&A1994}. Data 
from \citet{DraineApJ1984} and \citet{JamesMNRAS2002} are converted to 
$\kappa_\nu/\kappa_\mathrm{V}$ as in \citet{BianchiA&A1999} and 
\citet{AltonSub1999}. Points at 850$\mu$m have been slightly shifted in
$\lambda$ for ease of presentation.
}
\label{emco}
\end{figure}

The derived submm/mm emissivities are shown in Fig.~\ref{emco} together 
with estimates from literature. Emissivities derived from IRAS/COBE 
observations of diffuse Galactic dust \citep{BoulangerA&A1996,BianchiA&A1999} 
are within 10\% of the values predicted by the popular \citet{DraineApJ1984} 
model, which is shown in Fig.~\ref{emco} as a solid line. Dotted lines refer 
to the \citet{OssenkopfA&A1994} models for bare grains, grains with a thin ice 
coating and grains with thin ice after coagulation has proceeded for 10$^5$ 
years in a gas with density 10$^6$~g~cm$^{-3}$ (from bottom to top, 
respectively). 

\citet{JamesMNRAS2002} derived the emissivity from a sample of local galaxies 
observed with SCUBA, estimating the dust mass from metals.
Their value is compatible to that of \citet{DraineApJ1984}, indicating 
similar properties for diffuse dust in external galaxies and in the Milky Way.
Our 850~$\mu$m emissivity is higher than that of diffuse dust: 
a comparison with the models of \citet{OssenkopfA&A1994} seems to suggest that 
grains in Barnard~68 possess molecular ice mantles or have coagulated 
into fluffy aggregates. However, $\kappa_{850\mu\mathrm{m}}/\kappa_\mathrm{V}$ 
is within 2$\sigma$ of that for bare grains. In a very recent paper,
\citet{KramerA&A2002} measured $\kappa_{850\mu\mathrm{m}}/\kappa_\mathrm{V}$
for a molecular ridge and four embedded cores in the dark cloud IC5146.
Their radially averaged values of the four cores,
derived assuming $\beta$=2, range from $2.1\cdot 10^{-5}$ to $6.1\cdot
10^{-5}$ and are broadly compatible with our determination. At 1.2 mm,
our emissivity is similar to that of diffuse dust but lower than the value 
previously estimated by \citet{KramerA&A1998} on a IC5146 core.

The ratio $\kappa_{850\mu\mathrm{m}}/\kappa_{1.2\mathrm{mm}}$ we derive
implies  a value for $\beta$ larger than what expected from measurements 
on Galactic diffuse dust and on laboratory cosmic dust analogues 
\citep{MennellaApJ1998}. However, the error on the ratio is large and
$\beta=2$ is still compatible (within 1-$\sigma$) with our result.
Contamination of the submm and mm fluxes by the 
\element[][12]{CO}(3-2) and \element[][12]{CO}(2-1) lines
\citep{AveryApJ1987} was found to be negligible.

Clearly, more observations are needed to narrow down the errors in the 
determination of the emissivity. The analysis presented here has to be 
repeated on a large sample of objects of relatively simple morphology 
like Barnard~68. Only with a statistical sample will it be possible 
to study the variation of dust properties with the environment.

%__________________________________________________________________
\begin{acknowledgements}
We are grateful to J. Alves, R. Cesaroni, A.  Natta, C. Lada and C. Kramer 
(the referee) for useful comments and discussions.
\end{acknowledgements}

%__________________________________________________________________
%__________________________________________________________________
\bibliographystyle{aa}
\bibliography{../../DUST}

%__________________________________________________________________
\end{document}